\documentstyle[aps,preprint,floats,tighten,epsf]{revtex}

\begin{document}
\preprint{\vbox{\hbox{TRI--PP--00--08}
                \hbox{hep-lat/0003011}}}

\title{S and P-wave heavy-light mesons in lattice NRQCD}

\author{Randy Lewis}
\address{Department of Physics, University of Regina, Regina, SK, Canada 
         S4S 0A2}
\author{R. M. Woloshyn}
\address{TRIUMF, 4004 Wesbrook Mall, Vancouver, BC, Canada V6T 2A3}

\date{March 2000}
\maketitle

\begin{abstract}
The mass spectrum of S and P-wave mesons containing a single heavy
quark is computed in the quenched approximation, using NRQCD up to
third order in the inverse heavy quark mass expansion.
Previous results found third order contributions which are as large in
magnitude as the total second order contribution
for the charmed S-wave spin splitting.
The present work considers variations such as anisotropic
lattices, Landau link tadpole improvement, and a highly-improved light quark
action, and finds that the second order correction to the charmed
S-wave spin splitting is about 20\% of the leading order contribution,
while the third order correction is about 20\%(10\%) for $D^*-D$($D_s^*-D_s$).
Nonleading corrections are very small for the bottom meson spectrum, and are
statistically insignificant for the P-wave charmed masses.
The relative orderings among P-wave charmed and bottom mesons, 
and the sizes of the mass splittings, 
are discussed in light of experimental data and existing calculations.
\end{abstract}


\section{INTRODUCTION}

Calculations in QCD can be performed numerically on a discrete
space-time lattice, provided the lattice spacing is small enough to
accommodate all of the revelant physical distance scales.
In the presence of a heavy quark the lattice spacing must be
small relative to the inverse quark mass, resulting in large computational 
requirements, unless an appropriate ``effective theory'' is used.
In particular, for the case of a hadron containing exactly one
heavy quark, the dynamics can be expanded in powers of the inverse
heavy quark mass using the well-established technique of heavy quark 
effective theory\cite{HQET} or nonrelativistic QCD (NRQCD)\cite{NRQCD,NRQCD2}.
With the heavy quark expansion, the lattice spacing can be much coarser and 
the computational requirements are correspondingly smaller.

In the present work, two issues will be addressed within quenched lattice 
NRQCD.
Firstly, is the charm quark sufficiently heavy to permit the use of lattice 
NRQCD for charmed meson spectroscopy?  Secondly, what are the mass splittings
(magnitude and sign) between P-wave mesons containing a single 
heavy quark?

The first issue is clearly of interest due to the computational efficiency
of lattice NRQCD.  The heavy quark expansion is known to
work well for bottom mesons\cite{Swave,otherS}, but
previous research has demonstrated that $O(1/M^2)$ and $O(1/M^3)$
contributions can be comparable in magnitude for the S-wave spin splitting
of charmed mesons.\cite{Swave}  The present work provides an extension
of this investigation by considering new simulations that incorporate
a number of changes in method.  For example, Ref. \cite{Swave} used the
fourth root of an elementary plaquette to define the tadpole improvement
factor, $U_0$, whereas the present work uses the mean link in Landau gauge.
Since the Landau definition is advantageous in other contexts\cite{Landau},
including the velocity expansion for the charmonium spectrum of lattice 
NRQCD\cite{SFU}, it might be expected to improve the convergence of the 
heavy quark expansion for the charmed heavy-light spectrum as well.
Also, the light quark was described by the Sheikholeslami-Wohlert action in
Ref. \cite{Swave}, but the present work uses a D234 action\cite{D234} which
has smaller lattice spacing errors classically.
The simulations in the present work differ from those of Ref. \cite{Swave}
in various other ways as well, including
Dirichlet versus periodic boundary
conditions for light quark propagation, differing discretizations for
heavy quark propagation, and the introduction of an anisotropic lattice with
a smaller temporal lattice spacing than spatial spacing.
Despite these modifications, the present simulations produce a conclusion 
similar to that of Ref. \cite{Swave}: the $O(1/M^2)$ and $O(1/M^3)$ 
contributions are comparable in magnitude for the S-wave charmed spin 
splitting; in the present work, each is roughly 20\% of the $O(1/M)$ result
for $D^*-D$, though the $O(1/M^3)$ contribution is closer to 10\% for
$D_s^*-D_s$.

The second issue under discussion relates to the
spectrum of P-wave mesons containing a single heavy quark.
The relative orderings of the P-wave bottom mesons have only recently come
under direct experimental scrutiny\cite{Pexpt,CLEO,CDF,L3,ALEPH}, 
and the complete picture is not yet clear.  Meanwhile theoretical predictions
differ from one another even at a qualitative level.  The traditional
expectation of a hydrogen-like spectrum that arises from a number of 
model calculations\cite{Godfrey,Dai,Gupta,Orsland,Lahde} has been questioned
long ago by Schnitzer\cite{Schnitzer} and very recently by Isgur\cite{Isgur}
and by Ebert, Galkin and Faustov\cite{Ebert}.

The calculation is difficult within lattice QCD because the P-wave splittings
are not large in comparison to the typical scale of nonperturbative QCD, and
because of operator mixing for the pair of P-wave states having $J=1$.
Some previous attempts have been made\cite{Michael,Boyle,lat99,PUK,Hein}.
Unfortunately, the uncertainties are often substantial,
and results are not always as consistent with one another as 
might have been hoped.  The present work represents a further comment on this
situation.  
In particular, the $D_{s2}^*-D_{s0}^*$ mass splitting is 
found to be positive as in the traditional hydrogen-like ordering, and
to be substantially less than 100 MeV, while the $D_2^*-D_0^*$,
$B_{s2}^*-B_{s0}^*$ and $B_2^*-B_0^*$ splittings are even smaller.
These splittings are consistent with a number of model calculations,
but are somewhat smaller than the lattice NRQCD calculation of 
Ref. \cite{PUK}.

\section{ACTION}

The lattice action has three terms: gauge action, light quark action
and heavy quark action.
The entire action is classically and tadpole-improved with
the tadpole factors, $u_s$ and $u_t$, defined as the mean links in
Landau gauge in a spatial and temporal direction, respectively.

The gauge field action is
\begin{eqnarray}
S_G(U) &=& \frac{5\beta}{3}\left[
          \frac{1}{u_s^4\xi}
             \sum_{\rm ps}\left(1-\frac{1}{3}{\rm ReTr}U_{\rm ps}\right)
        - \frac{1}{20u_s^6\xi}
             \sum_{\rm rs}\left(1-\frac{1}{3}{\rm ReTr}U_{\rm rs}\right)
          \right.\nonumber \\
     && + \frac{\xi}{u_s^2u_t^2}
             \sum_{\rm pt}\left(1-\frac{1}{3}{\rm ReTr}U_{\rm pt}\right)
        - \frac{\xi}{20u_s^4u_t^2}
             \sum_{\rm rst}\left(1-\frac{1}{3}{\rm ReTr}U_{\rm rst}\right)
          \nonumber \\
     && \left.
        - \frac{\xi}{20u_s^2u_t^4}
             \sum_{\rm rts}\left(1-\frac{1}{3}{\rm ReTr}U_{\rm rts}\right)
          \right],
\end{eqnarray}
where $\xi \equiv a_s/a_t$.  The subscripts
``ps'' and ``rs'' denote spatial plaquettes and spatial planar 1$\times$2
rectangles respectively.  Plaquettes in the temporal-spatial planes are
denoted by ``pt'', while rectangles with the long side in a spatial(temporal)
direction are labeled by ``rst''(``rts'').
The leading classical errors of this action are quartic in lattice spacing.

For light quarks, a D234 action\cite{D234} is used with parameters set
to their tadpole-improved classical values.  Its leading classical errors are
cubic in lattice spacing.
\begin{eqnarray}
S_F(\bar{q},q;U) 
    &=& \frac{4\kappa}{3}\sum_{x,i}\left[\frac{1}{u_s\xi^2}D_{1i}(x)
        -\frac{1}{8u_s^2\xi^2}D_{2i}(x)\right]
      + \frac{4\kappa}{3}\sum_x\left[\frac{1}{u_t}D_{1t}(x)
        -\frac{1}{8u_t^2}D_{2t}(x)\right] \nonumber \\
  &&  + \frac{2\kappa}{3u_s^4\xi^2}\sum_{x,i<j}\bar\psi(x)\sigma_{ij}F_{ij}(x)
        \psi(x)
      + \frac{2\kappa}{3u_s^2u_t^2\xi}\sum_{x,i}\bar\psi(x)\sigma_{0i}F_{0i}(x)
        \psi(x) \nonumber \\
  &&  - \sum_x\bar\psi(x)\psi(x),
\end{eqnarray}
where
\begin{eqnarray}
D_{1i}(x) &=& \bar\psi(x)(1-\xi\gamma_i)U_i(x)\psi(x+\hat{i})
             +\bar\psi(x+\hat{i})(1+\xi\gamma_i)U_i^\dagger(x)\psi(x), \\
D_{1t}(x) &=& \bar\psi(x)(1-\gamma_4)U_4(x)\psi(x+\hat{t})
             +\bar\psi(x+\hat{t})(1+\gamma_4)U_4^\dagger(x)\psi(x), \\
D_{2i}(x) &=& \bar\psi(x)(1-\xi\gamma_i)U_i(x)U_i(x+\hat{i})\psi(x+2\hat{i})
             \nonumber \\
          &+& \bar\psi(x+2\hat{i})(1+\xi\gamma_i)U_i^\dagger(x+\hat{i})
              U_i^\dagger(x)\psi(x), \\
D_{2t}(x) &=& \bar\psi(x)(1-\gamma_4)U_4(x)U_4(x+\hat{t})\psi(x+2\hat{t})
             \nonumber \\
          &+& \bar\psi(x+2\hat{t})(1+\gamma_4)U_4^\dagger(x+\hat{t})
              U_4^\dagger(x)\psi(x), \\
gF_{\mu\nu}(x) &=& \frac{1}{2i}\left(\Omega_{\mu\nu}(x)-\Omega^\dagger_{\mu\nu}
                   (x)\right) - \frac{1}{3}{\rm Im}\left({\rm Tr}\Omega_{\mu
                   \nu}(x)\right), \\
\Omega_{\mu\nu} &=& \frac{-1}{4}\left[
   U_\mu(x)U_\nu(x+\hat\mu)U_\mu^\dagger(x+\hat\nu)U_\nu^\dagger(x) \right.
       \nonumber \\
       && ~~+U_\nu(x)U_\mu^\dagger(x-\hat\mu+\hat\nu)U_\nu^\dagger(x-\hat\mu)
       U_\mu(x-\hat\mu) \nonumber \\
       && ~~+U_\mu^\dagger(x-\hat\mu)U_\nu^\dagger(x-\hat\mu-\hat\nu)
       U_\mu(x-\hat\mu-\hat\nu)U_\nu(x-\hat\nu) \nonumber \\
       && ~~\left.
       +U_\nu^\dagger(x-\hat\nu)U_\mu(x-\hat\nu)U_\nu(x+\hat\mu-\hat\nu)
       U_\mu^\dagger(x) \right].
\end{eqnarray}

The heavy quark action is NRQCD\cite{NRQCD}, which is discretized to give
the following Green's function propagation:
\begin{eqnarray}\label{HQprop}
G_{\tau+1} &=& \left(1-\frac{a_tH_B}{2}\right)
    \left(1-\frac{a_tH_A}{2n}\right)^n\frac{U_4^\dagger}{u_t}
    \left(1-\frac{a_tH_A}{2n}\right)^n
    \left(1-\frac{a_tH_B}{2}\right) G_\tau,
\end{eqnarray}
with $n=5$ chosen for this work.
Separation of the Hamiltonian into two terms, $H=H_A+H_B$, is important
for ensuring stability of the discretization.  For example, recall the
discussion in Ref. \cite{Swave} of a large nonzero vacuum expectation value
for the term containing $c_{10}$ in the Hamiltonian (see Eq. (\ref{dH3})).
This issue will be discussed further in Sec. \ref{sec:Swave}.

The following Hamiltonian, written in terms of the bare heavy quark mass $M$,
is complete to $O(1/M^3)$ in the classical continuum limit\cite{Manohar}:
\begin{eqnarray}
H &=& H_0 + \delta{H}, \label{H} \\
H_0 &=& \frac{-\Delta^{(2)}}{2M}, \\
\delta{H} &=& \delta{H}^{(1)} + \delta{H}^{(2)} + \delta{H}^{(3)} + O(1/M^4) \\
\delta{H}^{(1)} &=& -\frac{c_4}{u_s^4}\frac{g}{2M}\mbox{{\boldmath$\sigma$}}
                    \cdot\tilde{\bf B} + c_5\frac{a_s^2\Delta^{(4)}}{24M}, \\
\delta{H}^{(2)} &=& \frac{c_2}{u_s^2u_t^2}\frac{ig}{8M^2}(\tilde{\bf \Delta}
                    \cdot\tilde{\bf E}-\tilde{\bf E}\cdot\tilde{\bf \Delta})
               -\frac{c_3}{u_s^2u_t^2}\frac{g}{8M^2}\mbox{{\boldmath$\sigma$}}
               \cdot(\tilde{\bf \Delta}\times\tilde{\bf E}-\tilde{\bf E}\times
           \tilde{\bf \Delta}) - c_6\frac{a_s(\Delta^{(2)})^2}{16n{\xi}M^2}, \\
\delta{H}^{(3)} &=& -c_1\frac{(\Delta^{(2)})^2}{8M^3}
                    -\frac{c_7}{u_s^4}\frac{g}{8M^3}\left\{\tilde\Delta^{(2)},
                        \mbox{{\boldmath$\sigma$}}\cdot\tilde{\bf B}\right\}
                    -\frac{c_9ig^2}{8M^3}\mbox{{\boldmath$\sigma$}}\cdot
                     \left(\frac{\tilde{\bf E}\times\tilde{\bf E}}{u_s^4u_t^4}
                    +\frac{\tilde{\bf B}\times\tilde{\bf B}}{u_s^8}\right) 
                     \nonumber \\
             && -\frac{c_{10}g^2}{8M^3}\left(\frac{\tilde{\bf E}^2}{u_s^4u_t^4}
                +\frac{\tilde{\bf B}^2}{u_s^8}\right)
                -c_{11}\frac{a_s^2(\Delta^{(2)})^3}{192n^2{\xi^2}M^3}.
              \label{dH3}
\end{eqnarray}
The coefficients of the Hamiltonian are chosen so the dimensionless
parameters, $c_i$, are unity at the classical level.
As will be discussed below, computations have been performed with the $c_i$ 
set to unity or zero in various combinations, including 
separate computations at $O(1/M)$, $O(1/M^2)$ and $O(1/M^3)$ to allow
discussions of convergence for the $1/M$ expansion.
Throughout this work, $H_0$ is always placed in $H_A$ of Eq. (\ref{HQprop})
and all of the remaining terms except the $c_{10}$ term are only placed 
in $H_B$.  The difference between having the $c_{10}$ term in $H_A$ or $H_B$ 
will be discussed explicitly, since it has the nonzero vacuum expectation 
value.

A tilde on any quantity indicates that the leading
discretization errors have been removed.  In particular,
\begin{eqnarray}
   \tilde{E}_i &=& \tilde{F}_{4i}, \\
   \tilde{B}_i &=& \frac{1}{2}\epsilon_{ijk}\tilde{F}_{jk},
\end{eqnarray}
where\cite{NRQCD2}
\begin{eqnarray}
   \tilde{F}_{\mu\nu}(x) &=& \frac{5}{6}F_{\mu\nu}(x) 
           - \frac{1}{6u_\mu^2}U_\mu(x)F_{\mu\nu}(x+\hat\mu)U_\mu^\dagger(x)
           - \frac{1}{6u_\mu^2}U_\mu^\dagger(x-\hat\mu)F_{\mu\nu}(x-\hat\mu)
             U_\mu(x-\hat\mu) \nonumber \\
       &&  - (\mu\leftrightarrow\nu).
\end{eqnarray}
The various spatial lattice derivatives are defined as follows:
\begin{eqnarray}
   a_s\Delta_iG(x) &=& \frac{1}{2u_s}[U_i(x)G(x+\hat\imath)
                    -U^\dagger_i(x-\hat\imath)G(x-\hat\imath)], \\
   a_s\Delta^{(+)}_iG(x) &=& \frac{U_i(x)}{u_s}G(x+\hat\imath) - G(x), \\
   a_s\Delta^{(-)}_iG(x) &=& G(x) -
              \frac{U^\dagger_i(x-\hat\imath)}{u_s}G(x-\hat\imath), \\
   a_s^2\Delta^{(2)}_iG(x) &=& \frac{U_i(x)}{u_s}G(x+\hat\imath) - 2G(x)
               +\frac{U^\dagger_i(x-\hat\imath)}{u_s}G(x-\hat\imath), \\
   \tilde\Delta_i &=& \Delta_i
                   - {a_s^2\over 6} \Delta^{(+)}_i\Delta_i\Delta^{(-)}_i, \\
   \Delta^{(2)} &=& \sum_i \Delta^{(2)}_i \label{Laplacian}, \\
   \tilde \Delta^{(2)} &=& \Delta^{(2)} - {a_s^2 \over 12} \Delta^{(4)}, \\
   \Delta^{(4)} &=& \sum_i \left( \Delta^{(2)}_i \right)^2.
\end{eqnarray}
This NRQCD action has quadratic classical lattice spacing errors.

\section{METHOD}\label{sec:method}

The data presented here come from 2000 gauge field configurations on
10$^3\times$30 lattices at $\beta=2.1$ with a bare aspect ratio of 
$\xi \equiv a_s/a_t=2$.
Two light quark masses are used, corresponding to $\kappa=0.23$ and 0.24.
Fixed time boundaries are used for the light quark propagators so they fit
naturally into a heavy-light meson, since the NRQCD heavy quark propagator
is also not periodic in the temporal direction.

A calculation of the string tension from these gauge field configurations
provides a determination of the renormalized anisotropy:
\begin{equation}
\xi \equiv a_s/a_t = 1.96(2).
\end{equation}

Using light quarks only, the lightest pseudoscalar and vector 
meson masses are easily obtained from local creation operators.
By linear interpolation and extrapolation in $1/\kappa$, the 
critical ($\kappa_c$) and strange ($\kappa_s$) hopping 
parameters and the temporal lattice spacing are found to be
\begin{eqnarray}
\kappa_c &=& 0.243025(41), \\
\kappa_s &=& \left\{\begin{tabular}{ll} 0.2344(11) & from $m_\phi$, \\
             0.2356(3) & from $m_K$, \end{tabular}\right. \label{kappa_s} \\
a_t &=& 0.1075(23){\rm fm~from~} m_\rho. \label{at(rho)}
\end{eqnarray}
For the hopping parameters used in explicit computations,
$\kappa=0.23$ and 0.24,
the ratio of pseudoscalar to vector meson masses is
``$m_\pi/m_\rho$'' = 0.815(3) and 0.517(8) respectively.
No exceptional configurations were encountered at these $\kappa$ values.
One might expect a systematic uncertainty on $a_t$ to account for
deviations from the linear relationship between $a_tm_\rho$ and
$1/\kappa$.  The uncertainty is presumably a few percent, but cannot
be estimated using only the two $\kappa$ values studied here.

A heavy-light meson is created by the following operator,
\begin{equation}
   \sum_{\vec x}Q^\dagger(\vec{x})\Omega(\vec{x})\Gamma(\vec{x})q(\vec{x}),
\end{equation}
where $\Omega(\vec{x})$ is given in Table~\ref{tab:operators} and the smearing
operator is
\begin{equation}
   \Gamma(\vec{x}) = [1+c_s\Delta^{(2)}(\vec{x})]^{n_s}.
\end{equation}
All plots shown here use $(c_s,n_s)=(0.15,10)$ at the source and a local sink.
The source is fixed at timestep 4, which is a distance $3a_t$ from the 
lattice boundary.
\begin{table}
\caption{Heavy-light meson creation operators.}\label{tab:operators}
\begin{tabular}{rl}
${}^{2S+1}L_J$ & $\Omega(\vec{x})$ \\
\hline
${}^1S_0$ & (0,$I$) \\
${}^3S_1$ & (0,$\sigma_i$) \\
${}^1P_1$ & (0,$\Delta_i$) \\
${}^3P_0$ & (0,$\sum_i\Delta_i\sigma_i$) \\
${}^3P_1$ & (0,$\Delta_i\sigma_j-\Delta_j\sigma_i$) \\
${}^3P_2$ & (0,$\Delta_i\sigma_i-\Delta_j\sigma_j$) or \\
          & (0,$\Delta_i\sigma_j+\Delta_j\sigma_i$), $i \neq j$
\end{tabular}
\vspace*{-5mm}
\end{table}

\begin{figure}[tbh]
\epsfxsize=380pt \epsfbox[30 325 498 725]{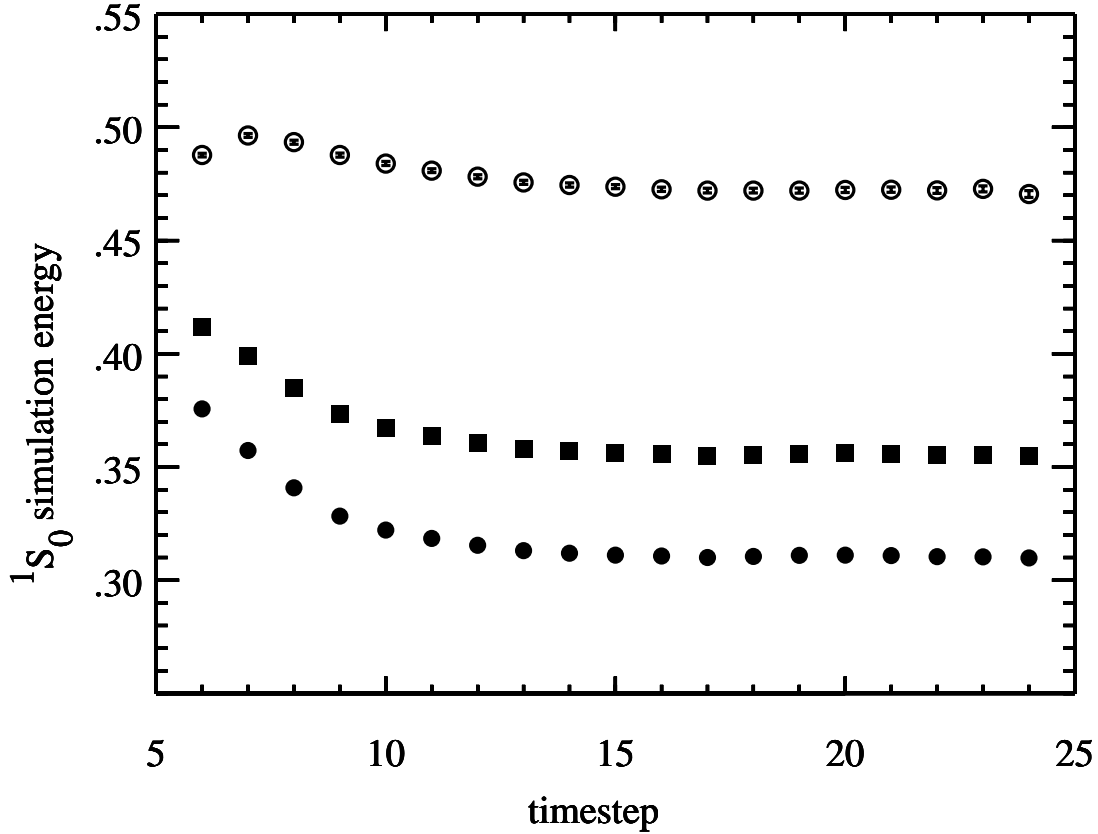}
\vspace{18pt}
\caption{Effective mass plots for the ground state heavy-light meson at rest
         with $\kappa=0.23$, and terms up to $O(1/(a_sM)^2)$ retained in the
         Hamiltonian.  Only $H_0$ is in $H_A$ of Eq. (\protect\ref{HQprop}).
         Solid circles, solid squares and open circles are for 
         $a_sM=1.2$, 1.5 and 5.0 respectively.  
         Statistical errors are smaller than the plotted symbols.
         }\label{fig:plateau}
\end{figure}
Because NRQCD is an expansion in the inverse {\it bare\/} heavy quark mass,
all meson mass differences can be obtained from correlation functions at
${\bf p} = {\bf 0}$, but the absolute meson mass itself cannot be obtained
directly. One way to 
determine the mass is to compute the change in energy when a meson is
boosted,
\begin{equation}\label{Mkin}
   E_{\bf p} - E_0 = \frac{\bf p^2}{2M_{\rm kin}}.
\end{equation}
This defines the kinetic mass, $M_{\rm kin}$, which is interpreted as the
meson's physical mass.  For the present work, $E_{\bf p}$ is computed
only for the ${}^1S_0$ state, with ${\bf p} = (0,0,2\pi/L_s)$ where 
$L_s=N_sa_s$ is the spatial extent of the lattice in physical units.
Solving for the kinetic mass gives
\begin{equation}
M_{\rm kin} = \frac{2\pi^2}{N_s^2\xi^2a_t[a_t(E_{\bf p}-E_0)]}.
\end{equation}
Some justification for Eq. (\ref{Mkin}) comes from consideration of the
next correction term, giving
\begin{equation}
   E_{\bf p} - E_0 = \frac{\bf p^2}{2M_{\rm kin}} 
                   - \frac{\bf p^4}{8M_{\rm kin}^3}.
\end{equation}
We have verified that the extra term shifts the kinetic mass by an
amount which is smaller than the uncertainties for every value of 
$M_{\rm kin}$ reported in this work.

In the case of S-wave mesons, a plateau containing ample timesteps 
is clearly evident in all effective mass plots; some examples are shown in
Fig. \ref{fig:plateau}.
A more detailed discussion of the P-wave plateaus is
deferred to Sec. \ref{sec:Pwave}.
In this paper,
the plateau region for each mass is defined by the maximum value of 
\begin{equation}
Q \equiv \frac{\Gamma(N/2-1,\chi^2/2)}{\Gamma(N/2-1,0)}
\end{equation}
where
\begin{equation}
\Gamma(a,x) = \int_x^\infty{\rm d}t\,t^{a-1}\exp(-t),
\end{equation}
and $N$ is the number of timesteps in the proposed plateau region.
$\chi^2$ is obtained from a single exponential fit to each meson correlation
function.  A correlated fit is done with the covariance matrix inverted by
singular value decomposition.
Statistical uncertainties are obtained from the analysis of 5000 bootstrap
ensembles.  All plateaus are ended at timestep 22(20) for simulations with 
$\kappa=0.23(0.24)$, except for an infinitely-heavy quark due to excessive
noise at these larger times.  
The examples in Table \ref{tab:Q} demonstrate the quality of the fits.
\begin{table}
\caption{Examples of plateaus defined by maximization of $Q$.
Notice that they all have the desired feature that $Q \geq 0.1$.
Insensitivity of the fit parameters to the precise plateau boundaries
has been verified.}\label{tab:Q}
\begin{tabular}{cccccccc}
& ${}^1S_0$ & ${}^1S_0({\bf p})$ & ${}^3S_1$ &
${}^1P_1$ & ${}^3P_0$ & ${}^3P_1$ & ${}^3P_2$ \\
\hline
\multicolumn{8}{l}{$a_sM = 1.5$, $\kappa = 0.23$} \\
($t_{\rm min},t_{\rm max}$) & (15,22) & (16,22) & (17,22) & 
(13,22) & (12,22) & (14,22) & (12,22) \\
Q & 0.45 & 0.13 & 0.27 & 0.60 & 0.21 & 0.97 & 0.92 \\
\hline
\multicolumn{8}{l}{$a_sM = 1.5$, $\kappa = 0.24$} \\
($t_{\rm min},t_{\rm max}$) & (15,20) & (16,20) & (15,20) &
(12,20) & (11,20) & (14,20) & (13,20) \\
Q & 0.22 & 0.14 & 0.61 & 0.69 & 0.26 & 0.86 & 0.43 \\
\hline
\multicolumn{8}{l}{$a_sM = 6$, $\kappa = 0.23$} \\
($t_{\rm min},t_{\rm max}$) & (17,22) & (16,22) & (16,22) &
(13,22) & (12,22) & (13,22) & (12,22) \\
Q & 0.88 & 0.29 & 0.67 & 0.15 & 0.69 & 0.65 & 0.50 \\
\hline
\multicolumn{8}{l}{$a_sM = 6$, $\kappa = 0.24$} \\
($t_{\rm min},t_{\rm max}$) & (16,20) & (14,20) & (16,20) &
(13,20) & (12,20) & (13,20) & (12,20) \\
Q & 0.51 & 0.16 & 0.77 & 0.35 & 0.10 & 0.12 & 0.93
\end{tabular}
\vspace*{-5mm}
\end{table}

\section{S-WAVE SPECTRUM}\label{sec:Swave}

\begin{figure}[tbh]
\epsfxsize=380pt \epsfbox[30 325 498 725]{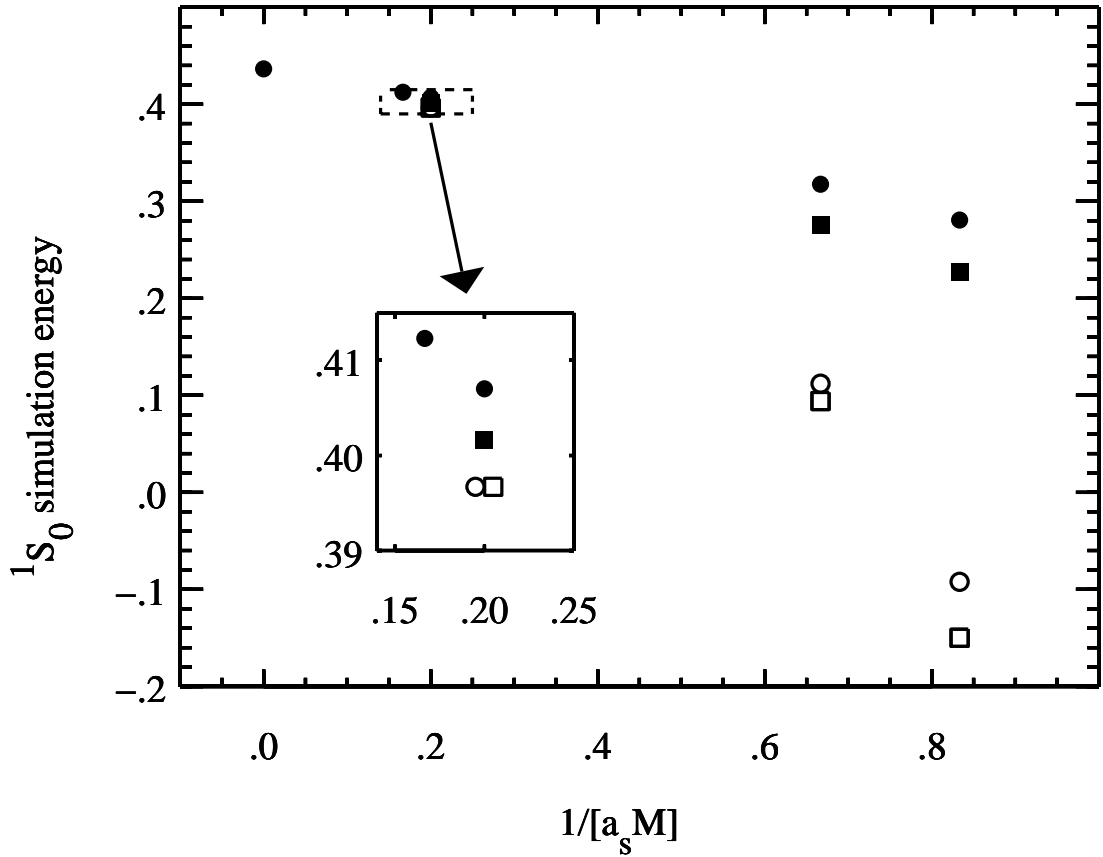}
\vspace{18pt}
\caption{The simulation energy of a ground state heavy-light meson at rest.
         Results are displayed from terms up to $O(1/(a_sM)^k)$, with 
         $k=1,2,3$.  
         $\kappa$ is fixed at 0.24, and $M$ is the bare heavy quark mass.
         Solid circles are $k=1$, solid squares are $k=2$, 
         open circles(squares) are $k=3$ with the $c_{10}$ term in $H_B(H_A)$
         of Eq. (\protect\ref{HQprop}).
         Statistical errors are smaller than the plotted symbols.
         }\label{fig:sim}
\end{figure}
Calculations were performed for $a_sM$ = 1.2, 1.5, 5, 6 and $\infty$, where $M$ 
represents the bare heavy quark mass in the NRQCD action.
Figure \ref{fig:sim} shows the simulation energy of the ground state as a
function of $a_sM$ for $\kappa = 0.24$.  The huge $O(1/M^3)$ effect at
smaller $M$ values is due to the vacuum expectation value of the $c_{10}$ term.
This large correction to the unphysical simulation energy does not discredit 
the convergence of NRQCD, but special care must be taken to ensure that the
large vacuum value is incorporated into the heavy quark propagation 
appropriately.  In particular, previous work\cite{Swave} has shown the linear
approximation to be insufficient for the $c_{10}$ term, which contains the
vacuum expectation value, in computations of
the S-wave spin splitting in the charm region.
Fig. \ref{fig:sim} explicitly shows the error introduced by placing the
$c_{10}$ term in $H_B$ rather than $H_A$ in Eq. (\ref{HQprop}).

It will also be noted from Fig. \ref{fig:sim} that the simulation energy
is negative for $a_sM=1.2$ at third order in NRQCD.
Of course the absolute energy scale is unphysical in NRQCD due to omission
of the large heavy quark mass term from the leading order action, and
physical quantities (i.e. mass differences) are independent of the absolute 
energy scale.
If the simulation energy were {\it large\/} and negative, it might signal a
poor $1/M$ expansion and/or a
problem for the discretization of heavy quark propagation, but a small
negative result presents no problem.

Table \ref{tab:kineticmass} shows a calculation of the kinetic mass, and
indicates that the bottom quark requires $a_sM \approx 5.5$.  The charm
quark seems to want $1.2 < a_sM < 1.5$, although the data suggest that
the $1/M$ expansion might not be converging in this region.
It is interesting to notice that the vacuum expectation value does not affect
the calculation of this observable significantly.

\begin{table*}
\caption{The kinetic energy of a ${}^1S_0$ heavy-light meson.
         $H_A$ and $H_B$ are defined by Eq. (\protect\ref{HQprop}) 
         and $c_{10}$ by Eq. (\protect\ref{dH3}).  Except for $a_sM=6$, the 
         results in physical units are computed from $O(1/M^2)$ data,
         using the lattice spacing from Eq. (\protect\ref{at(rho)}) to set the
         physical length scale.  Only statistical uncertainties are shown.
         }\label{tab:kineticmass}
\begin{tabular}{lllllll}
         & $a_sM$ & \multicolumn{4}{c}{$a_tE(\vec{p})-a_tE(\vec{0})$} 
         & $M_{\rm kin}$ [GeV] \\
\cline{3-6}
     & & $O(1/M)$ & $O(1/M^2)$ & \multicolumn{2}{c}{$O(1/M^3)$} & \\
\cline{5-6}
     & &          &            & $c_{10}$ in $H_B$ & $c_{10}$ in $H_A$ & \\
\hline
$\kappa$=0.23 
     & 1.2 & 0.0487(10) & 0.0515(10) & 0.0530(10) & 0.0582(39)$^*$ & 1.83(6) \\
     & 1.5 & 0.0425(9)  & 0.0444(9)  & 0.0459(9)  & 0.0519(25)$^*$ & 2.12(8) \\
     & 5.0 & 0.0185(10) & 0.0188(10) & 0.0188(10) & 0.0239(26)$^*$ & 5.0(3) \\
     & 6.0 & 0.0162(10) &            &            &              & 5.8(4) \\
\hline
$\kappa$=0.24 
     & 1.2 & 0.0496(25) & 0.0523(24) & 0.0549(24) & 0.0566(27)   & 1.80(10) \\
     & 1.5 & 0.0430(24) & 0.0449(22) & 0.0469(21) & 0.0471(23)   & 2.10(12) \\
     & 5.0 & 0.0191(17) & 0.0194(16) & 0.0194(16) & 0.0194(16)   & 4.9(4) \\
     & 6.0 & 0.0180(22) &            &            &              & 5.2(7)
\end{tabular}
${}^*$These computations use only 200 configurations.
\end{table*}

A stronger statement about convergence comes from the splitting
between the spin-singlet and spin-triplet S-wave states, since the
uncertainties are smaller.  Table \ref{tab:Ssplit} suggests a nice
convergence in the bottom region, but no guarantee of convergence for charm.
It will be noted that an incorrect treatment of the vacuum expectation value
(i.e. putting the $c_{10}$ term into $H_B$)
can actually lead to a small $O(1/M^3)$ contribution, but this is incidental.
\begin{table*}
\caption{The ${}^3S_1-{}^1S_0$ mass splitting.
         $H_A$ and $H_B$ are defined by Eq. (\protect\ref{HQprop}) 
         and $c_{10}$ by Eq. (\protect\ref{dH3}).  Except for $a_sM=6$, the 
         results in physical units are computed from $O(1/M^2)$ data,
         using the lattice spacing from Eq. (\protect\ref{at(rho)}) to set the
         physical length scale.  Only statistical uncertainties are shown.
         }\label{tab:Ssplit}
\begin{tabular}{lllllll}
         & $a_sM$ & \multicolumn{4}{c}{$a_tM({}^3S_1)-a_tM({}^1S_0)$} 
         & $M({}^3S_1)-M({}^1S_0)$ \\
\cline{3-6}
     & & $O(1/M)$ & $O(1/M^2)$ & \multicolumn{2}{c}{$O(1/M^3)$} & [MeV] \\
\cline{5-6}
     & &          &            & $c_{10}$ in $H_B$ & $c_{10}$ in $H_A$ & \\
\hline
$\kappa$=0.23 
     & 1.2 & 0.0441(6)  & 0.0527(7)  & 0.0508(7) & 0.0566(25)$^*$ & 96.6(24) \\
     & 1.5 & 0.0385(6)  & 0.0447(6)  & 0.0447(6) & 0.0452(19)$^*$ & 81.9(21) \\
     & 5.0 & 0.0155(4)  & 0.0165(4)  & 0.0165(4) & 0.0159(8)$^*$  & 30.2(10) \\
     & 6.0 & 0.0136(5)  &            &            &              & 24.9(11) \\
\hline
$\kappa$=0.24 
     & 1.2 & 0.0483(11) & 0.0589(12) & 0.0577(11) & 0.0677(15)   & 107.9(32) \\
     & 1.5 & 0.0418(9)  & 0.0494(11) & 0.0502(10) & 0.0554(11)   & 90.5(28) \\
     & 5.0 & 0.0147(13) & 0.0157(13) & 0.0158(13) & 0.0158(13)   & 28.8(25) \\
     & 6.0 & 0.0136(7)  &            &            &              & 24.9(14)
\end{tabular}
${}^*$These computations use only 200 configurations.
\end{table*}

Tables \ref{tab:kineticmass} and \ref{tab:Ssplit} correctly accommodate the 
vacuum expectation value of the $c_{10}$ term by placing it in $H_A$.  
In Ref. \cite{Swave}, this method was found to give the same numerical
results, within statistical uncertainties, for the S-wave kinetic mass 
and spin splitting as was obtained by computing the vacuum expectation 
value directly and subtracting it from the Hamiltonian.
For the present calculation, a similar check was performed: 
the vacuum expectation value was computed from 400 of the gauge field 
configurations, and the S-wave kinetic mass and spin splitting were computed
from 200 configurations using the Hamiltonian with the vacuum value
explicitly removed.  As expected, the results agree within statistics with
Tables \ref{tab:kineticmass} and \ref{tab:Ssplit} when the $c_{10}$ term 
is in $H_A$.

Interpolating these data so that $M_{\rm kin}$ is the physical mass of a 
bottom meson,
one finds a spin splitting which is only about 55\% of the experimental value.
This is typical of quenched lattice calculations (see for eg.
Refs. \cite{Swave,Boyle,PUK}).  Even an unquenched NRQCD calculation
did not reproduce the experimental $B^*-B$ splitting\cite{unquenched}  
so perhaps the tadpole-improved {\it classical\/} values, which were used
for the coefficients $c_i$ in the NRQCD Hamiltonian, account for the residual
discrepancy.

The data reported in Ref. \cite{Swave} display a $1/M$ expansion
for charmed mesons in lattice NRQCD in which 
$O(1/M^3)$ terms were as significant as $O(1/M^2)$ terms, so convergence
of the expansion could not be assured.
In that work, it was hoped that a more convergent expansion might 
be obtained via changes in the lattice method.  In particular, replacement
of the average plaquette tadpole factor by the mean link in Landau gauge
was suggested to hold some promise.\cite{Landau,SFU}
The present work has made this 
modification plus a number of others including: 
a more aggressively-improved light quark action,
asymmetric lattices with temporal spacing reduced by a factor of two,
Dirichlet temporal boundaries for light quarks rather than periodic ones,
smeared meson sources rather than local sources,
and a symmetric dependence on $H_B$ in Eq. (\ref{HQprop}).
Despite these changes, convergence of the $1/M$ expansion remains 
uncomfirmed for S-wave charmed mesons.
It is possible that a study of terms beyond $O(1/M^3)$ would reveal
that the series really is well-behaved, but this remains unexplored.

\section{P-WAVE SPECTRUM}\label{sec:Pwave}

\begin{figure}[tbh]
\epsfxsize=380pt \epsfbox[30 325 498 725]{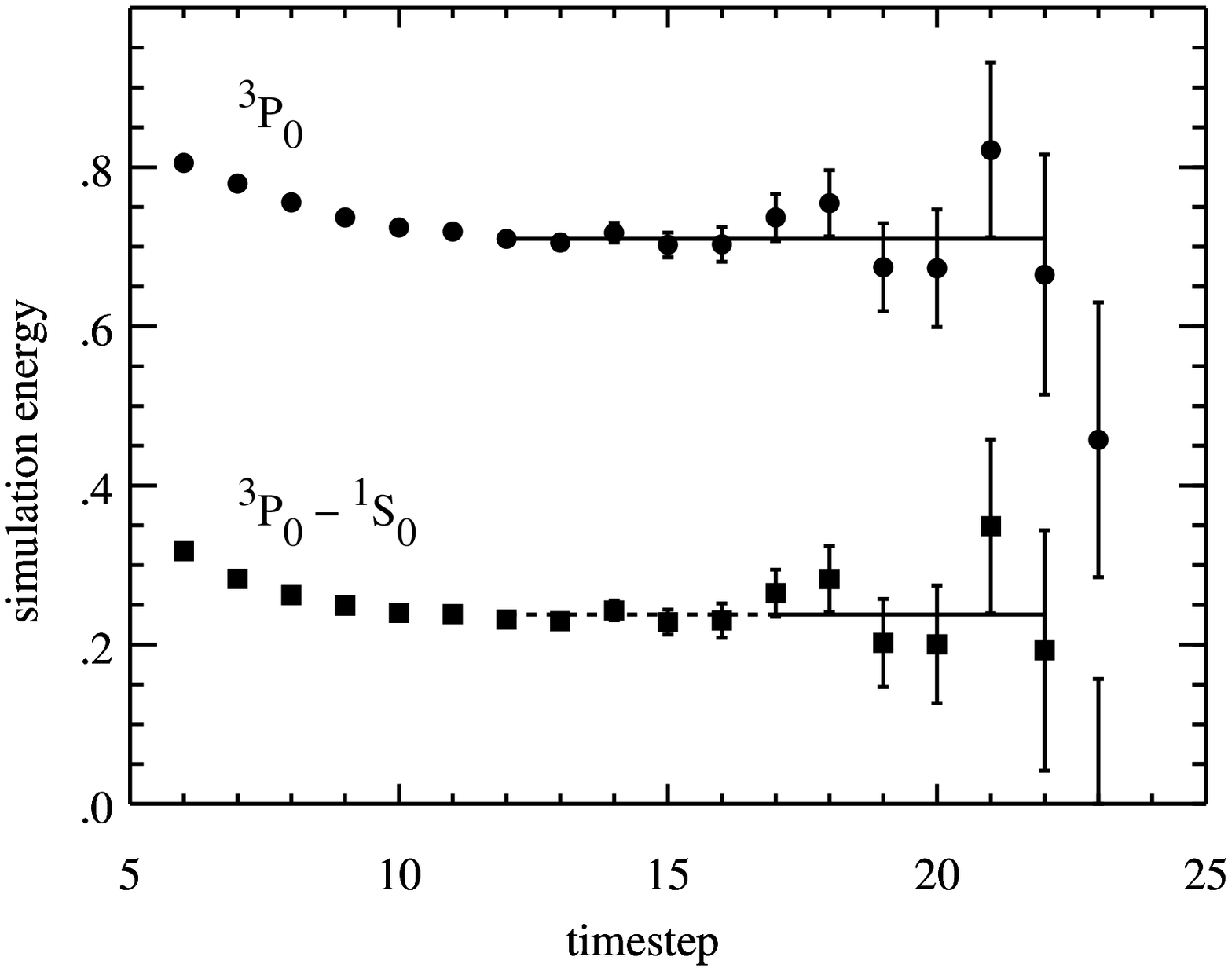}
\vspace{18pt}
\caption{The simulation energy of a ${}^3P_0$ heavy-light meson, for
         $\kappa=0.23$ and $a_sM=5.0$.  The Hamiltonian contains all terms 
         up to and including $O(1/(a_sM)^2)$.
         The splitting between the ${}^3P_0$ and ${}^1S_0$ is also shown.
         }\label{fig:spsim}
\end{figure}

Each of the four P-wave operators from Table \ref{tab:operators} leads to a
visually-identifiable plateau; an example is shown in Fig. \ref{fig:spsim}.
The method of maximum $Q$, discussed in
Sec. \ref{sec:method}, can be used to define precise plateau boundaries
and the resultant ${}^3P_0-{}^1S_0$ splitting is shown in Table \ref{tab:SP0}.
In contrast to the S-wave splitting discussed in the previous section, the
$1/M$ corrections to the ${}^3P_0-{}^1S_0$ splitting are not large
relative to the statistical uncertainties, even in the charm region.
The splittings given in Table \ref{tab:SP0} are consistent
with the available experimental data, as will be discussed below.
\begin{table*}
\caption{The ${}^3P_0-{}^1S_0$ mass splitting.
         $H_A$ and $H_B$ are defined by Eq. (\protect\ref{HQprop}) 
         and $c_{10}$ by Eq. (\protect\ref{dH3}).  
         Except for $a_sM=6$ and $a_sM=\infty$, the 
         results in physical units are computed from $O(1/M^2)$ data,
         using the lattice spacing from Eq. (\protect\ref{at(rho)}) to set the
         physical length scale.  Only statistical uncertainties are shown.
         }\label{tab:SP0}
\begin{tabular}{lllllll}
         & $a_sM$ &
           \multicolumn{4}{c}{$a_tM({}^3P_0)-a_tM({}^1S_0)$} 
         & $M({}^3P_0)-M({}^1S_0)$ \\
\cline{3-6}
     & & $O(1/M)$ & $O(1/M^2)$ & \multicolumn{2}{c}{$O(1/M^3)$} & [MeV] \\
\cline{5-6}
     & &          &            & $c_{10}$ in $H_B$ & $c_{10}$ in $H_A$ & \\
\hline
$\kappa$=0.23 
     & 1.2 & 0.275(5) & 0.271(4) & 0.276(4) & 0.277(16)$^*$ & 497(13) \\
     & 1.5 & 0.266(5) & 0.263(4) & 0.267(4) & 0.268(10)$^*$ & 482(13) \\
     & 5.0 & 0.238(5) & 0.238(4) & 0.238(4) & 0.252(11)$^*$ & 436(12) \\
     & 6.0 & 0.236(5) &          &          &             & 432(13) \\
  &$\infty$& 0.225(5) &          &          &             & 412(13) \\
\hline
$\kappa$=0.24 
     & 1.2 & 0.311(6) & 0.311(6) & 0.315(6) & 0.321(7)    & 570(16) \\
     & 1.5 & 0.300(6) & 0.301(6) & 0.304(6) & 0.306(6)    & 552(16) \\
     & 5.0 & 0.256(8) & 0.256(7) & 0.256(7) & 0.256(7)    & 469(16) \\
     & 6.0 & 0.254(8) &          &          &             & 465(18) \\
  &$\infty$& 0.210(20) &         &          &             & 385(38) \\
\end{tabular}
${}^*$These computations use only 200 configurations.
\end{table*}

Tables \ref{tab:SP1SP2} and \ref{tab:PP1PP2} contain lattice results 
for splittings which involve the other P-wave mesons.
In the charm region, the $\kappa = 0.23$ data produce a ${}^3P_2$ meson 
which is heavier than the ${}^3P_0$ meson, but at $\kappa = 0.24$ the
${}^3P_2 - {}^3P_0$ splitting is consistent with zero.  The magnitudes
of the splittings decrease in the bottom region, as expected.

\begin{figure}[tbh]
\epsfxsize=380pt \epsfbox[30 60 498 625]{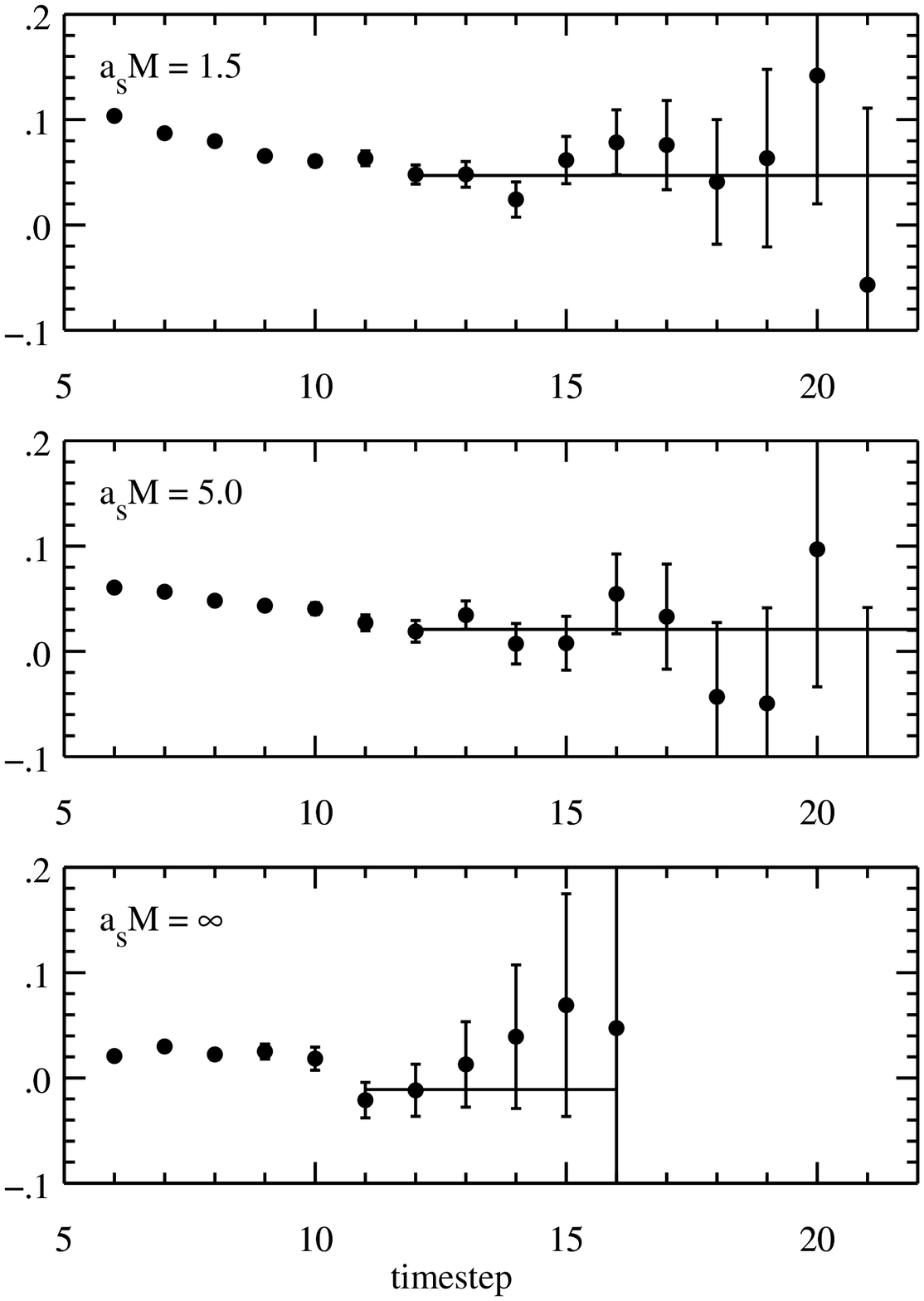}
\caption{Effective mass plots for the ${}^3P_2-{}^3P_0$ mass splitting
         with $\kappa=0.23$ and $a_sM=1.5$, 5.0 and $\infty$.
         The plateau region and value is also shown.
         }\label{fig:pdiff}
\end{figure}

A comparison of Tables \ref{tab:SP0},
\ref{tab:SP1SP2} and \ref{tab:PP1PP2}
plus the effective mass plots in Fig. \ref{fig:pdiff}
provide some indication of the systematic uncertainty which arises from the
choice of plateau region.
In particular, it will be noted that the effective mass plots are monotonically
decreasing near the source, so if a plateau region is chosen too near the
source, it will produce a mass splitting which is too large.
Thus one arrives at an upper bound for the ${}^3P_2-{}^3P_0$ splitting, 
as presented in Ref. \cite{lat99}.

\begin{table*}
\caption{The ${}^1P_1-{}^1S_0$, ${}^3P_1-{}^1S_0$ and ${}^3P_2-{}^1S_0$ 
         mass splittings.
         Data at $a_sM=6$ use the Hamiltonian up to $O(1/M)$ and 
         data at $a_sM<6$ use the Hamiltonian up to $O(1/M^2)$.
         Only statistical uncertainties are shown.
         }\label{tab:SP1SP2}
\begin{tabular}{lllll}
   & $a_sM$ & \multicolumn{3}{c}{$a_tM(X)-a_tM({}^1S_0)$} \\
\cline{3-5}
   &        & $X={}^1P_1$ & $X={}^3P_1$ & $X={}^3P_2$ \\
\hline
$\kappa$=0.23 
     & 1.2 & 0.288(7) & 0.298(9) & 0.325(7) \\
     & 1.5 & 0.279(7) & 0.285(8) & 0.311(7) \\
     & 5.0 & 0.236(8) & 0.246(7) & 0.259(7) \\
     & 6.0 & 0.240(7) & 0.243(8) & 0.253(7) \\
  &$\infty$& 0.181(26)& 0.226(10) & 0.214(11)\\
\hline
$\kappa$=0.24 
     & 1.2 & 0.295(13) & 0.309(20) & 0.327(18) \\
     & 1.5 & 0.295(10) & 0.294(19) & 0.315(17) \\
     & 5.0 & 0.232(15) & 0.250(14) & 0.270(13) \\
     & 6.0 & 0.225(16) & 0.246(15) & 0.265(13) \\
  &$\infty$& 0.200(32) & 0.257(61) & 0.219(17) \\
\end{tabular}
\end{table*}

\begin{table*}
\caption{The ${}^1P_1-{}^3P_0$, ${}^3P_1-{}^3P_0$ and ${}^3P_2-{}^3P_0$ 
         mass splittings.
         Data at $a_sM=6$ use the Hamiltonian up to $O(1/M)$ and 
         data at $a_sM<6$ use the Hamiltonian up to $O(1/M^2)$.
         Only statistical uncertainties are shown.
         }\label{tab:PP1PP2}
\begin{tabular}{lllll}
   & $a_sM$ & \multicolumn{3}{c}{$a_tM(X)-a_tM({}^3P_0)$} \\
\cline{3-5}
   &        & $X={}^1P_1$ & $X={}^3P_1$ & $X={}^3P_2$ \\
\hline
$\kappa$=0.23 
     & 1.2 &  0.017(7)  & 0.027(9) & 0.054(7) \\
     & 1.5 &  0.015(7)  & 0.021(8) & 0.047(7) \\
     & 5.0 & -0.002(8)  & 0.008(6) & 0.021(7) \\
     & 6.0 &  0.004(7)  & 0.007(7) & 0.017(7) \\
  &$\infty$& -0.044(26) & 0.001(9) & -0.011(11) \\
\hline
$\kappa$=0.24 
     & 1.2 & -0.016(14) & -0.002(20) & 0.016(19) \\
     & 1.5 & -0.005(10) & -0.007(19) & 0.014(17) \\
     & 5.0 & -0.024(16) & -0.006(13) & 0.014(14) \\
     & 6.0 & -0.029(17) & -0.008(14) & 0.011(15) \\
  &$\infty$& -0.010(32) &  0.047(21) & 0.009(27) \\
\end{tabular}
\end{table*}

The correlation functions constructed using ${}^1P_1$ and ${}^3P_1$ operators
contain some combination of the physical $J=1$ mesons. Both calculations
should lead to the same (lighter) physical mass at large Euclidean times
if both operators have a substantial overlap with the less massive $J=1$ state.
In principle, the masses of the physical states can be obtained using
the ${}^1P_1$/${}^3P_1$ operator basis 
by also calculating the mixing matrix
elements, but the effect was too small to be observed in these data.
In practice, for Euclidean times which can be used in our lattice 
simulation, no significant energy difference is observed between the 
${}^1P_1$ and ${}^3P_1$ channels.

Predictions for the physical mesons are displayed alongside experimental 
data in Table \ref{tab:Pexpt}.  The $D_s-D$ and $B_s-B$ mass differences 
depend only mildly on $O(1/M^2)$ and $O(1/M^3)$ terms, and they are close
to experiment.  In contrast, the S-wave spin splittings are significantly
smaller than experiment (as is typical of quenched lattice 
QCD\cite{Swave,Boyle,PUK}).  Furthermore, the $O(1/M^2)$ and $O(1/M^3)$
corrections to $D^*-D$ are each about 20\% of the leading order result.
The results in Table \ref{tab:Pexpt} use the value of $\kappa_s$
determined from $m_K$ in Eq.~(\ref{kappa_s}).  Use of the other
determination in Eq.~(\ref{kappa_s}) shifts the central values of
the $D_s^*-D_s$ and $B_s^*-B_s$
mass splittings by 1 MeV or less, and shifts the splittings among
P-waves ($D_s^{**}$ and $B_s^{**}$) by 7 MeV or less.

A detailed comparison of P-wave results would be somewhat premature, since
the experimental data are rather incomplete and often rely on theoretical
models for input, while the lattice calculation is quenched and lacks a
firm connection between the physical mesons and the $J=1$ operators.
Nevertheless, Table \ref{tab:Pexpt} shows a general consistency between
the experimental and computed P-wave masses.

\begin{table*}
\caption{The heavy-light spectrum compared to experiment.
         The hadron naming scheme of Ref. \protect\cite{PDG99} is
         followed\protect\cite{DPRIME}.  The charm and bottom masses are
         fixed to the experimental $D_s$ and $B_s$ masses (giving
         $a_sM_c = 1.37(8)$ and $a_sM_b = 5.8(^{+13}_{-6})$).  The first
         uncertainty on lattice data combines the statistical error with
         the uncertainty of $a_t$ from Eq. (\ref{at(rho)}).  The second
         uncertainty on lattice data corresponds to the errors in $a_sM_c$
         and $a_sM_b$.  The value of $\kappa_s$ determined from $m_K$ in 
         Eq.~(\ref{kappa_s}) has been used.
         }\label{tab:Pexpt}
\begin{tabular}{cccccc}
\multicolumn{3}{c}{charmed meson masses [MeV]} 
& \multicolumn{3}{c}{bottom meson masses [MeV]} \\
\cline{1-3}\cline{4-6}
               & experiment/Ref. & lattice & 
               & experiment/Ref. & lattice \\
\hline
$D_s$          & 1969(1)/\cite{PDG99}          & input             & 
$B_s$          & 5369(2)/\cite{PDG99}          & input             \\
$D_s^*-D_s$    & 144/\cite{PDG99}              & $O(1/M)$:79(2)(3) & 
$B_s^*-B_s$    & 47(3)/\cite{PDG99}            & $26(1)(^{+3}_{-6})$     \\
               &                               & $O(1/M^2)$:94(2)(5) & 
               &                               &                   \\
               &                               & $O(1/M^3)$:103(3)(6) & 
               &                               &                   \\
$D_{s0}^*-D_s$ &                               & 530(13)(5)        & 
$B_{s0}^*-B_s$ &                               & $451(15)(^{+3}_{-6})$   \\
$D_{s1}'-D_s$  &                               & ${}^1P_1$:531(16)(2) & 
$B_{s1}'-B_s$  &                            & ${}^1P_1$:$425(24)(^{+2}_{-5})$\\
$D_{s1}-D_s$   & 566(1)/\cite{PDG99}           & ${}^3P_1$:542(22)(8) &
$B_{s1}-B_s$   &                            & ${}^3P_1$:$449(22)(^{+5}_{-9})$\\
$D_{s2}^*-D_s$ & 604(2)/\cite{PDG99}           & 585(19)(6)        & 
$B_{s2}^*-B_s$ &                               & $478(21)(^{+6}_{-12})$  \\
               &                               &                   & 
($B_s^{**}-B_s$) & 484(15)/\cite{PDG99}$\dagger$ &                   \\
\hline
$D_s-D$        & 99,104/\cite{PDG99}           & $O(1/M)$:105(2)(1)& 
$B_s-B$        & 90(2)/\cite{PDG99}            & $92(3)(^{+1}_{-0})$     \\
               &                               & $O(1/M^2)$:107(3)(1)& 
               &                               &                   \\
               &                               & $O(1/M^3)$:112(4)(1)& 
               &                               &                   \\
$D^*-D$        & 141,142/\cite{PDG99}          & $O(1/M)$:84(3)(3) & 
$B^*-B$        & 46/\cite{PDG99}               & $25(2)(^{+3}_{-4})$     \\
               &                               & $O(1/M^2)$:101(3)(5) & 
               &                               &                   \\
               &                               & $O(1/M^3)$:117(4)(6) & 
               &                               &                   \\
$D_0^*-D$      &                               & 579(15)(5)        & 
$B_0^*-B$      &                               & $475(19)(^{+4}_{-6})$   \\
$D_1'-D$       & 596(53)/\protect\cite{CLEO}   & ${}^1P_1$:548(20)(2) & 
$B_1'-B$       & 391(16)/\cite{L3}*       & ${}^1P_1$:$407(38)(^{+11}_{-24})$\\
$D_1-D$        & 558(2)/\cite{CLEO}            & ${}^3P_1$:557(32)(8) & 
$B_1-B$        & 431(20)/\cite{CDF}*       & ${}^3P_1$:$453(33)(^{+5}_{-10})$\\
$D_2^*-D$      & 595(2)/\cite{PDG99}           & 588(30)(6)        & 
$B_2^*-B$      & 489(8)/\cite{L3}*             & $493(29)(^{+5}_{-13})$  \\
               &                               &                   & 
               & 460(13)/\cite{ALEPH}*         &                   \\
               &                               &                   & 
($B^{**}-B$) & 418(9)/\cite{PDG99}$\dagger$  & 
\end{tabular}
$\dagger$Experimental signal is a sum over resonances with differing 
momenta $J=0,1,2$.

*Theoretical estimates for some of the mass splittings have
been used as input.
\end{table*}

It is instructive to compare lattice P-wave masses to the 
predictions of models, such as quark models\cite{Godfrey,Isgur,Ebert},
a Bethe-Salpeter study\cite{Dai}, a chromodynamic potential model\cite{Gupta},
a bag model\cite{Orsland} and a Blankenbecler-Sugar approach\cite{Lahde}.
Many of these present results for $J=1$ directly in the ${}^1P_1/{}^3P_1$
basis which simplifies the comparison to lattice data, and the quark models
may in general be more closely related to the quenched approximation than to 
experiment.

The models of Refs. \cite{Godfrey,Dai,Gupta,Lahde} predict the traditional
hydrogen-like ordering of P-waves, where the ${}^3P_0$ is the lightest meson,
the ${}^3P_2$ is the heaviest, and the $P_1$ states lie in between.  
The authors of Ref. \cite{Orsland} find,
from lightest to heaviest, ${}^3P_0$, $P_1(3/2)$, ${}^3P_2$, $P_1(1/2)$, where
the arguments represent the angular momentum of the light
degrees of freedom.  Ref. \cite{Isgur} predicts a dramatic inversion 
where the ${}^3P_0$ is heavier than the ${}^3P_2$ by 100(150) MeV for 
$D^{**}$($B^{**}$) mesons, but the $P_1$ states are the absolute
lightest and heaviest.
Ref. \cite{Ebert} claims very small splittings (tens of MeV) where the
$P(1/2)$ and $P(3/2)$ doublets overlap to produce different orderings for
the $D$, $D_s$ and $B$ systems (the $B_s$ is ordered like the $D$).

\begin{figure}[tbh]
\epsfxsize=380pt \epsfbox[30 0 498 725]{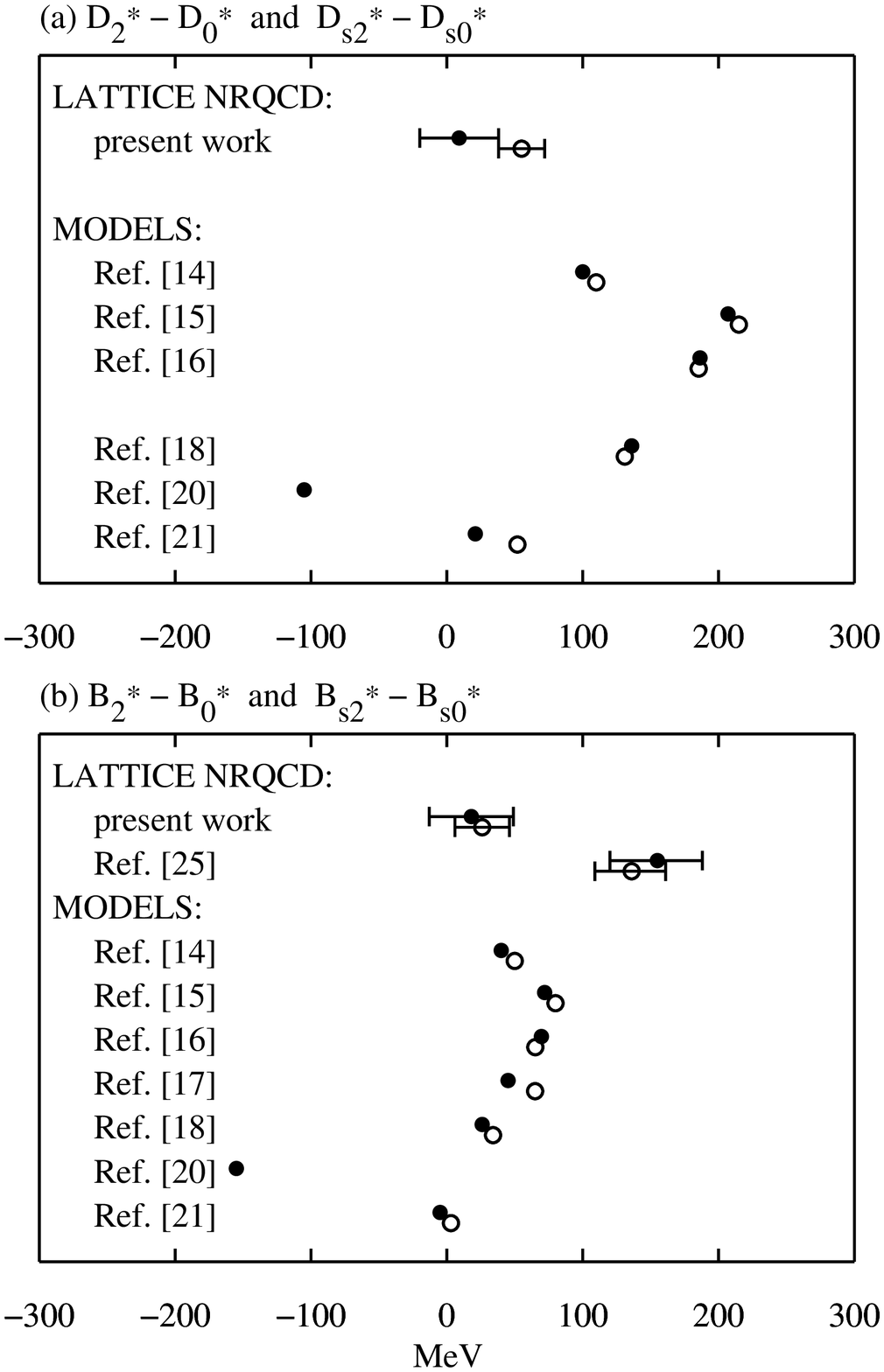}
\caption{The $D_2^*-D_0^*$ and $B_2^*-B_0^*$ splittings (solid symbols)
         and the $D_{s2}^*-D_{s0}^*$ and $B_{s2}^*-B_{s0}^*$ splittings 
         (open symbols) from lattice QCD and various model calculations.
         }\label{fig:models}
\end{figure}

\begin{figure}[tbh]
\epsfxsize=380pt \epsfbox[30 0 498 725]{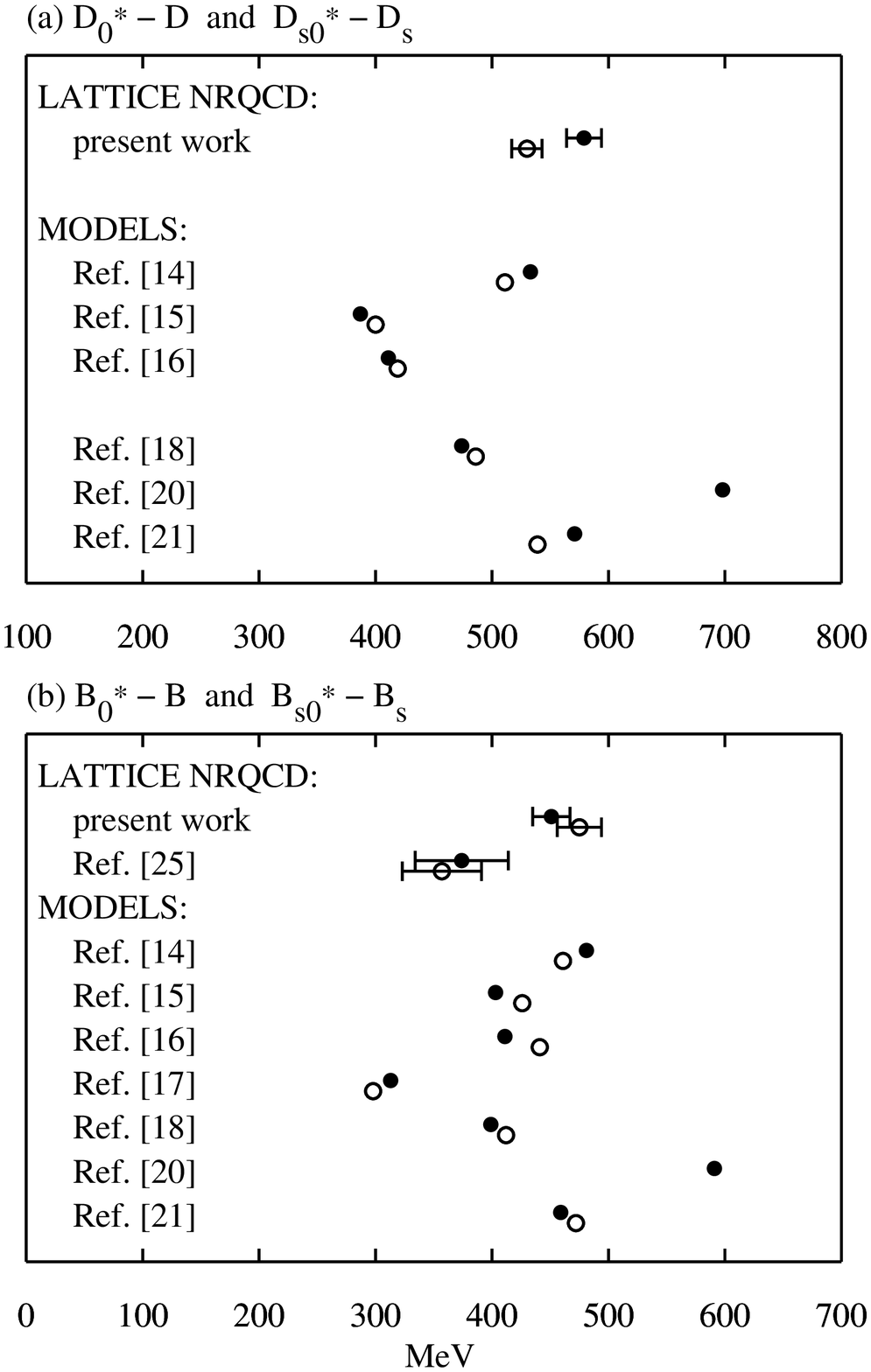}
\caption{The $D_0^*-D$ and $B_0^*-B$ splittings (solid symbols)
         and the $D_{s0}^*-D_s$ and $B_{s0}^*-B_s$ splittings 
         (open symbols) from lattice QCD and various model calculations.
         }\label{fig:modelP0}
\end{figure}

Fig. \ref{fig:models} shows the $D_2^*-D_0^*$ and $B_2^*-B_0^*$ splittings
from lattice QCD and from the models just mentioned.
Despite the range of model predictions, it is evident that
some general consistency exists between our results and a number of the
models.  Notice in particular that our results are numerically distinct 
from the large inversion of Ref. \cite{Isgur}.
Fig. \ref{fig:modelP0} compares the ${}^3P_0-{}^1S_0$ splittings as 
obtained from lattice QCD and the models.

Of special importance is the comparison with Ref. \cite{PUK}, where the 
spectrum was also computed from quenched lattice NRQCD.
The discrepancy between the two lattice calculations is
exemplified by Fig. \ref{fig:models}.
While there are many differences in method between the two computations,
it is difficult to identify a compelling reason for the disagreement.
Fig. \ref{fig:pdiff} indicates that our data must satisfy 
$M(B_2^*)-M(B_0^*) < 100$ MeV for any chosen plateau region, and this is
not consistent with Ref. \cite{PUK}.  It is hoped that further lattice efforts
will improve this situation.  Unfortunately, a recent lattice NRQCD study
of the heavy-light meson spectrum\cite{Hein} has statistical uncertainties 
which are too large to resolve the discrepancy between our results
and those of Ref. \cite{PUK}.

\section{CONCLUSIONS}

The masses of S and P-wave heavy-light mesons have been calculated in
the quenched approximation, using
lattice NRQCD for the heavy quark and a highly-improved action for the
light degrees of freedom.  Calculations at first, second and third order 
in the heavy quark mass expansion were used as a test of convergence, and
it was concluded that the $O(1/M^2)$ and $O(1/M^3)$ contributions to
the $D^*-D$ splitting are both near 20\%.  The $D_s^*-D_s$ splitting
receives a 20\% correction at $O(1/M^2)$ and a 10\% correction at $O(1/M^3)$.
These results might represent a convergent $1/M$ expansion if terms beyond 
$O(1/M^3)$ are decreasing appropriately, but convergence cannot be ascertained 
from the present study.  This same conclusion was reached in Ref. \cite{Swave} 
by a computational method which differed from the present 
one in some details; most notably, Ref. \cite{Swave} used the average 
plaquette tadpole definition whereas the present work uses the mean link
in Landau gauge.

No convergence problem is found for P-wave charmed masses, and the
P-wave spectrum for both charmed and bottom mesons is predicted.
The ${}^3P_2$ is heavier than the ${}^3P_0$ and $P_1$ states in the 
$D_s^{**}$ system, with $M(D_{s2}^*)-M(D_{s0}^*)=55 \pm 17$ MeV.
For the $D^{**}$, $B^{**}$ and $B_s^{**}$ systems the $P_2-P_0$
splittings are smaller and consistent with zero.
These conclusions agree with a number of model calculations and are 
compatible with the available experimental data.

\acknowledgments

The authors thank Howard Trottier for a critical reading of the manuscript,
and R.L. thanks Niranjan Venugopal for useful discussions.
This work was supported in part by the Natural Sciences and Engineering
Research Council of Canada.


\begin{references}
\bibitem{DPRIME} A $'$ is used to distinguish between the two physical $J=1$
    states. The $'$ superscript will be used for the state which in the heavy
    quark spin symmetry classification has light degrees of freedom with 
    angular momentum of $j_l=1/2$ and which is expected to have a large width.
    The unprimed $J=1$ state corresponds to $j_l=3/2$ and is expected to be
    narrow.
\end{references}
\end{document}